\documentclass[preprint,12pt]{elsarticle}
 
\usepackage{etoolbox}
\makeatletter
\def\ps@pprintTitle{%
	\let\@oddhead\@empty
	\let\@evenhead\@empty
	\def\@oddfoot{\reset@font\hfil\thepage\hfil}
	\let\@evenfoot\@oddfoot
}
\makeatother
\usepackage{bm}
\usepackage{float}
\usepackage{graphics}
\usepackage{amsmath, amsthm, amssymb}
\usepackage{natbib}
\usepackage{caption}
\usepackage[colorlinks=true]{hyperref}
\usepackage{lscape}
\usepackage{float}
\usepackage[utf8]{inputenc}
\usepackage{amsthm}

\usepackage{enumerate}
\usepackage{mathrsfs}
\usepackage{placeins}
\usepackage[inner=4cm,outer=2cm]{geometry}
\usepackage{graphicx}
\usepackage{amsfonts}
\usepackage{verbatim}
\usepackage{amssymb}
\usepackage{amsthm,multirow}
\usepackage{orcidlink}

\newtheorem{thm}{Theorem}[section]
\newtheorem{lem}{Lemma}[section]

\newtheorem{prop}{Proposition}[section]

\theoremstyle{definition}
\newtheorem{defn}{Definition}[section]
\newtheorem{ex}{Example}[section]

\theoremstyle{remark}
\newtheorem{rem}{Remark}[section]

\theoremstyle{properties}

\theoremstyle{Examples}

\numberwithin{equation}{section}

\usepackage[mathscr]{euscript}
\usepackage{geometry} 
\usepackage{graphicx,lscape} 
\usepackage{booktabs} 
\usepackage{bigstrut}
\usepackage{array}
\usepackage{paralist} 
\usepackage{verbatim}
\usepackage{subfig} 
\biboptions{comma,round,authoryear}
\begin{document}
\begin{frontmatter}
	\title{\textbf{Fractional Cumulative Residual Inaccuracy \\in the Quantile Framework and its Applications}}
	\author{Iona Ann Sebastian \orcidlink{0009-0000-7640-2215}\corref{cor1}}
    \author{S. M. Sunoj \orcidlink{0000-0002-6227-1506}}
	\ead{ionaann99@gmail.com, smsunoj@cusat.ac.in}
	\cortext[cor1]{Corresponding author}

	\address{Department of Statistics\\Cochin University of Science and Technology\\Cochin 682 022, Kerala, INDIA}

	\begin{abstract}
	Fractional cumulative residual inaccuracy (FCRI) measure allows to determine regions of discrepancy between systems, depending on their respective fractional and chaotic map parameters. Most of the theoretical results and applications related to the FCRI of the lifetime random variable are based on the distribution function approach. However, there are situations in which the distribution function may not be available in explicit form but has a closed-form quantile function (QF), an alternative method of representing a probability distribution. Motivated by these, the present study is devoted to introduce a quantile-based FCRI and study its various properties. We also deal with non-parametric estimation of quantile-based FCRI and examine its validity using simulation studies and illustrate its usefulness in measuring the discrepancy between chaotic systems and in measuring the discrepancy in two different time regimes using  Nifty 50 dataset. 
	\end{abstract}
	

\end{frontmatter}

%......................................................
\section{Introduction}
Information theory entails a diverse array of applications in different areas including statistics, physics, economics and engineering. Since 1948, researchers have done various studies on the generalization of Shannon's entropy. Following this, the idea of uncertainty, as developed in communication theory, is generalized to inaccuracy. We often encounter two types of errors because of insufficient information in the experimental results and the other because of misspecification of the model. To consider these errors, \cite{kerridge1961inaccuracy} proposed an inaccuracy measure. Let $f(\cdot),g(\cdot)$ represent the probability density functions (PDFs), and $\bar{F}(\cdot), \bar{G}(\cdot)$  the respective survival functions (SFs) corresponding to absolutely continuous non-negative random variables (rvs) $X$ and $Y$. If $f(\cdot)$ is the actual probability density function and $g(\cdot)$ is the assigned density function, then the inaccuracy measure of $X$ and $Y$ is given by (see \cite{kerridge1961inaccuracy})
\begin{equation}\label{krd in}
\mathcal{I}(X,Y)=-\int_{0}^{\infty}f(x)\log g(x)dx=-E_{f}(\log g(X)),   
\end{equation}
where the expectation is computed using the density function $f(\cdot)$. Clearly, \eqref{krd in} becomes Shannon differential entropy when $f(\cdot) = g(\cdot)$. For more properties of inaccuracy and associated measures, one can refer to \cite{kayal2017generalized}, \cite{ghosh2018generalized}, \cite{psarrakos2018residual}, \cite{di2019past}, \cite{hashempour2024new}, \cite{kharazmi2024fisher} and the references therein.\\

Since the SF exists in situations where
 the PDF does not, the information measures based on SFs are of greater importance. In this regard, the information measures based on cumulative distribution functions (CDF)
 have attracted significant attention  because they are inherently stable, as the CDF is generally more regular and exhibits less sensitivity to fluctuations compared to PDFs. After replacing PDFs with cumulative distribution functions (CDFs) in \eqref{krd in}, \cite{kumar2015dynamic}
introduced a measure that is an extension of cumulative past entropy known as the cumulative past inaccuracy (CPI) measure, given by (see also \cite{kundu2016cumulative})
\begin{equation}\label{cum past in}
\mathcal{PI}(X,Y)=-\int_{0}^{\infty}F(x)\log G(x)dx.   
\end{equation}
In the same way, the cumulative residual inaccuracy (CRI) is expressed as:
\begin{equation}\label{cum residual in}
\mathcal{RI}(X,Y)=-\int_{0}^{\infty}\bar{F}(x)\log \bar{G}(x)dx.   
\end{equation}
Several properties, such as non-negativity, monotonicity under transformations,  interpretability in reliability and risk settings, pertain to these cumulative inaccuracy measures.\\

To generalize these measures within the framework of fractional calculus, the notion of fractional cumulative residual inaccuracy (FCRI), which is an extension of the fractional cumulative residual entropy, was proposed by \cite{kharazmi2024fractional}, given by
\begin{equation}\label{fcri}
\mathcal{RI_\alpha}(X,Y)=-\int_{0}^{\infty} \bar{F}(x)[-\log \bar{G}(x)]^\alpha dx, 0 \leq \alpha \leq 1.
\end{equation}
The potential applications of this measure are highlighted in various fields such as statistics, information theory, image processing and machine learning. Clearly, in \eqref{fcri} when $\alpha=1$, the FCRI tends to cumulative residual inaccuracy (CRI). The parameter $\alpha$ characterizes fractals, whereas the fractal dimension is another indicator of complexity. The FCRI measure detects areas of divergence between systems by leveraging their fractional and chaotic map parameters. \\

Although abundant research is available for different uncertainty measures, a study of the FCRI utilizing quantile function (QF) does not seem to have been undertaken. Many authors have shown that the QF defined by
\begin{equation*}
	Q(u) = \inf \{x|F(x)\geq u\}, \; 0 \leq u \leq 1
\end{equation*}
is an efficient and equivalent alternative to the distribution function in modelling and analysis of statistical data (see \cite{gilchrist2000statistical}, \cite{nair2009quantile}). There are several characteristics of QFs that are not carried out
by distribution functions. For instance, two QFs are added together to form another QF. Since QF is less affected by outliers and offers a straightforward analysis using limited information, it is frequently more convenient.
For some recent studies on QF, 
Its properties and usefulness in the identification of models we refer to  \cite{vineshkumar2021inferring}, \cite{varkey2023review}, \cite{aswin2023reliability}, \cite{kayal2024quantile},  and the references therein. Also, many QFs used in applied works such as various forms of lambda distributions (\cite{gilchrist2000statistical}),  power-Pareto distribution (\cite{gilchrist2000statistical}, \cite{hankin2006new}), Govindarajulu distribution (\cite{nair2012modelling}) etc. do not have tractable distribution functions but have an explicit form of quantile functions.
This makes it difficult to use \eqref{fcri} to statistically study the characteristics of $\mathcal{RI}_{\alpha}(X,Y)$ for these distributions. Thus, the formulation
of the definition and properties of FCRI based on QFs is needed. The quantile-based FCRI (Q-FCRI) has several pros. First, unlike \eqref{fcri}, the suggested measure is much easier to compute when the distribution functions are intractable, while the QFs are of simple forms. Moreover, quantile functions have certain characteristics that probability distributions do not possess. 
Applying these characteristics yields some new findings and improved understanding of the measure that are not possible to obtain by the traditional method.\\

The layout of the paper is arranged as follows. Following the current introduction, we introduce the quantile-based fractional cumulative residual inaccuracy (Q-FCRI) and examine its various properties in Section 2. Followed by Section 3, where we suggest a non-parametric estimator for computing Q-FCRI. In Section 4, we executed simulation studies for evaluating the performance of the estimator given in Section 3.  Finally, in Sections 5 and 6, we illustrate a potential application of Q-FCRI in measuring the discrepancy between chaotic systems such as Chebyshev's and logistic maps and measuring the discrepancy in different financial time regimes using the Nifty 50 data set and concluding remarks of the study.

\section{Quantile-based FCRI}
Consider two absolutely continuous non-negative rvs, 
$X$ and $Y$ having CDFs $F(\cdot)$, $G(\cdot)$ and QFs $Q_X(\cdot)$, $Q_Y(\cdot)$. Further, $F(Q_X(u))=u$, $0<u<1$ and on diffrentiating $F(Q_X(u))=u$, we obtain
\begin{equation*}
q_X(u)f(Q_X(u))=1, 0<u<1.
\end{equation*}
Here, $q_X(u)$ denotes the quantile density function (QDF) and  $f(Q_X(u))$ denotes the density quantile function corresponding to $F$. Also, $Q_3(\cdot)=Q_Y^{-1}(Q_X(\cdot))=G(F^{-1}(\cdot))$ denote the QF of $F(G^{-1}(\cdot))$ and $q_3(u)=\frac{d}{du}Q_3(u)$ be the QDF of $Q_3(u)$. Thus, the quantile-based FCRI is defined as: 

\begin{defn}
\textnormal The quantile-based FCRI (Q-FCRI) corresponding to two non-negative absolutely continuous rvs $X$ and $Y$ with quantile fuctions $Q_X(\cdot)$ and $Q_Y(\cdot)$ is defined as
\begin{eqnarray}\label{qfcri}
\mathcal{RI_\alpha^{Q}}(X,Y)&=& \int_{0}^{1}\bar{F}(Q_X(p))(-\log(\bar{G}(Q_X(p))))^\alpha q_X(p)dp\\ 
{}&=&\int_{0}^{1}(1-p)(-\log(1-Q_3(p)))^\alpha q_X(p)dp.\nonumber
\end{eqnarray}	
\end{defn}
Using the expansion of the Maclurian series (\cite{spiegel1953some})
\begin{equation*}
-\log(1-x)=\sum_{i=1}^{n} \frac{x^i}{i}, 0<x<1,
\end{equation*}
a representation for the Q-FCRI measure in \eqref{qfcri}
 is given by
 \begin{equation*}
     \mathcal{RI_\alpha^Q}(X,Y)=\int_{0}^{1}(1-p)\bigg(\sum_{i=1}^{\infty}\frac{Q^i_3(p)}{i}\bigg)^\alpha q_X(p)dp.
\end{equation*}
When the distribution of $X$ is equal to that of $Y$, $i.e.,$ $Q_X(\cdot)=Q_Y(\cdot)$, \eqref{qfcri} reduces to the quantile-based fractional cumulative residual entropy (Q-FCRE) given as
\begin{equation}\label{QFCRE}
    \mathcal{E_\alpha^Q}(X)=\int_{0}^{1}(1-p)(-\log(1-p))^\alpha q_X(p)dp.
\end{equation}
It is clear that when $\alpha=1$, $\mathcal{RI_\alpha^Q}(X)$ becomes quantile-based cumulative residual inaccuracy (Q-CRI). The hazard quantile function (HQF) and mean residual quantile function (MRQF) serve as key tools for describing the physical behaviour of failure patterns.. Using the HQF, \eqref{qfcri} can be expressed as
\begin{equation}
\mathcal{RI_\alpha^Q}(X,Y)=\int_{0}^{1}[H_{XQ}(p)]^{-1}(-\log(1-Q_3(p)))^\alpha dp. \nonumber
\end{equation}
where $H_{XQ}(u)=[q_X(u)(1-u)]^{-1}$, for all $u \hspace{0.1cm}\epsilon  \hspace{0.1cm} (0,1)$. Also, MRQF uniquely determines the QDF given by (see \cite{nair2013quantile})
\begin{equation*}
q_{X}(u)=\frac{M_{X}(u)-(1-u)M'_{X}(u)}{(1-u)}.
\end{equation*}
where, $M_{X}(u)$ denotes the MRQF corresponding to $X$.
Thus, we can represent \eqref{qfcri} in terms of MRQF as
\begin{equation*}
\mathcal{RI_\alpha^Q}(X,Y)=\int_{0}^{1}(M_{X}(p)-(1-p)M_{X}'(p))(-\log(1-Q_3(p)))^\alpha dp.
\end{equation*}

\begin{center}
\begin{table}[H]\label{table 1}
\small
\caption{Some examples of Q-FCRI}
\begin{tabular}{  c c c c }
  \hline
$Q_X(u)$  & $Q_Y(u)$ & $\alpha$ & $\mathcal{RI_\alpha^Q}(X,Y)$\\ \hline
  $2u-u^2$  & $u$  & 0.25 & 0.546\\
 (special case of Govindarajulu)  & (uniform) &  0.5 & 0.482\\
   
    &  &  0.75 & 0.452\\
     &  &  1 & 0.609\\
     \hline
  $-\frac{1}{\lambda_1}\log(1-u)$ &   $-\frac{1}{\lambda_2}\log(1-u)$ & $0\leq \alpha \leq 1$ & $\frac{\lambda_2^\alpha}{\lambda_1^{\alpha+1}}\Gamma(\alpha+1)$\\
  \hline
 $r_1[1-(1-u)^{\frac{1}{c_1}}]$ & $r_2[1-(1-u)^{\frac{1}{c_2}}]$ & 0.5 & $\frac{c_2^{0.5} \sqrt{\pi}}{2c_1^{1.5} \big(\frac{1}{c_1}+1\big)^\frac{3}{2}}$\\
 (rescaled beta, $r_1=1$, $c_1>0$) & (rescaled beta, $r_2=1$, $c_2>0$) & 0.75
 & $\frac{c_2^{0.75} 3\Gamma(3/4)}{4c_1^{1.75} \big(\frac{1}{c_1}+1\big)^\frac{7}{4}}$\\
 \hline
 $Q_1(u)$ & $Q_1(1-(1-u)^\frac{1}{\theta})$ & $0 \leq \alpha \leq 1$ & $\theta^\alpha \mathcal{E}_{\alpha}^Q(X_1)$\\
 (Cox Proportional Hazard (PH) model) & & &\\
 \hline
$\frac{1}{(a+b)}\log\big(\frac{a+bu}{a(1+u)}\big)$ & $-\frac{1}{\lambda_2}\log(1-u)$ & 0.25 & $0.039 \lambda_2^{0.25}$\\
(linear hazard rate, a=1, b=2) & (exponential, $\lambda_2>0$) & 0.5 & $0.018\lambda_2^{0.5} $\\
& & 0.75 & $0.0089 \lambda_2^{0.75}$\\
\hline
%$\frac{1}{\lambda_1}+\frac{1}{\lambda_2}(u^{\lambda_3}-(1-u)^{\lambda_4})$ & $u$ & 0.5 & $-0.54+0.057i$\\
%(generalized lambda distribution, & (uniform) & & \\
% $\lambda_1=\lambda_2=\lambda_3=1$, $\lambda_4=0.2$ ) & & &\\
%\hline
\end{tabular}
\label{Table 1}
\end{table}
\end{center}
\begin{rem}
A particular case of Govindarajulu presented in Table 1, does not possess an explicit distribution function but has a QF. Therefore, in these cases, the survival function-based inaccuracy tool proposed by \cite{kharazmi2024fractional} is inapplicable, whereas an alternative tool to compute the same is the quantile-based inaccuracy given in \eqref{qfcri}.
\end{rem}

Some general properties of Q-FCRI with corresponding proofs are listed below:-
\begin{enumerate}
    \item Non-negativity: $\mathcal{RI_\alpha^Q}(X,Y) \geq 0$.
    \begin{proof}
    \begin{equation*}
    \mathcal{RI_\alpha^Q}(X,Y)=\int_{0}^{1}(1-p)(-\log(1-Q_3(p)))^\alpha q_X(p)dp.
    \end{equation*}
    Here, $Q_3(p)=Q_Y^{-1}(Q_X(p))=G(Q_X(p))$ implies $\log(1-Q_3(p))$ is negative. Hence, $(-\log(1-Q_3(p)))$ is positive. Therefore, the integrand $(1-p)(-\log(1-Q_3(p)))^\alpha q_X(p)$ is positive $\Rightarrow$ $\mathcal{RI_\alpha^Q}(X,Y) \geq 0.$ 
    \end{proof} 
    \item Linearity:- For $X'=aX+b$ and $Y'=cY+b$, with $a,c>0$ and $b\geq0$, it follows that 
    \begin{equation*}
        \mathcal{RI_\alpha^Q}(X',Y')=a \int_{0}^{1}(1-p)(-\log(1-Q_3(p)))^\alpha q_X(p)dp=a \mathcal{RI_\alpha^Q}(X,Y).
    \end{equation*}
    \begin{proof}
    \begin{eqnarray*}
    \mathcal{RI_\alpha}(X',Y')&=&\int_{0}^{\infty} \bar{F}'(x)(-\log \bar{G}'(x))^\alpha dx\\
    {}&=&\int_{b}^{\infty} \bar{F}'\bigg(\frac{x-b}{a}\bigg)\bigg(-\log \bar{G}'\bigg(\frac{x-b}{a}\bigg)\bigg)^\alpha dx, x \geq b.
    \end{eqnarray*}
    Now, on putting $z=\frac{x-b}{a}$,
    \begin{equation}\label{lin}
    \mathcal{RI_\alpha}(X',Y')=a\int_{0}^{\infty} \bar{F}(z)(-\log \bar{G}(z))^\alpha dz.
    \end{equation}
    The quantile-based form of \eqref{lin} is obtained as;
    \begin{eqnarray*}
\mathcal{RI_\alpha^Q}(X',Y')&=&a\int_{0}^{1}(1-p)(-\log(1-Q_3(p)))^\alpha q_X(p) dp\\
 {}&=& a \mathcal{RI_\alpha^Q}(X,Y).
    \end{eqnarray*}
\end{proof}
\item Joint convexity:- If $0 \leq \lambda \leq 1$ and $0 \leq \alpha  \leq 1$, it follows that
\begin{equation}\label{con}
\mathcal{RI_\alpha^Q}(\lambda X+(1-\lambda)Y, \lambda X'+(1-\lambda)Y') \leq \lambda \mathcal{RI_\alpha^Q}(X,X')+(1-\lambda) \mathcal{RI_\alpha^Q}(Y,Y').
\end{equation}
\begin{proof}
Firstly, we need to show that $\mathcal{RI_\alpha^Q}$ is a convex function, and thereafter we will prove the result using Jensen's inequality.

By applying the chain rule,
\begin{equation}\label{convex}
\frac{\partial}{\partial \alpha}\mathcal{RI_\alpha^Q}(X,Y)=\int_{0}^{1}(1-p)(-\log(1-Q_3(p)))^\alpha\log(-\log(1-Q_3(p)))q_X(p)dp.
\end{equation}
Again on differentiating \eqref{convex} with respect to $\alpha$,
\begin{equation}\label{convex2}
\frac{\partial^2}{\partial\alpha^2}\mathcal{RI_\alpha^Q}(X,Y)=\int_{0}^{1}(1-p)(-\log(1-Q_3(p)))^\alpha(\log(-\log(1-Q_3(p))))^2q_X(p)dp.
\end{equation}
Clearly, \eqref{convex2} is non-negative. Hence, $\mathcal{RI_\alpha^Q}(X,Y)$ is a convex function of $\alpha$. Thus, on applying Jensen's inequality, we can yield $\eqref{con}$.
\end{proof}

\item Antisymmetric:- $\mathcal{RI_\alpha^Q}(X,Y) \neq \mathcal{RI_\alpha^Q}(Y,X)$.  
%\begin{proof}
%According to the definition of FCRI, we have
%$$\mathcal{RI_\alpha}(X_2,X_1)=\int_{0}^{\infty} \bar{F}_2(x)(-\log \bar{F}_1(x))^\alpha dx$$. 
%From the definition of Q-FCRE we have
%\begin{equation*}
%\mathcal{RI_\alpha^Q}(X_1,X_2)=\int_{0}^{1}(1-p)(-\log(1-Q_3(p)))^\alpha q_1(p) dp
%\end{equation*}
%and,
%\begin{equation*}
%\mathcal{RI_\alpha^Q}(X_2,X_1)=\int_{0}^{1}(1-Q_3(p))(-\log(1-p))^\alpha q_1(p)dp
%\end{equation*}
%Clearly, %$\mathcal{RI_\alpha^Q}(X_1,X_2) \neq \mathcal{RI_\alpha^Q}(X_2,X_1)$. 
%\end{proof}
Therefore, when comparing two lifetime distributions in reliability modeling, it is plausible to analyze $\mathcal{RI_\alpha^Q}(Y,X)$ in addition to $\mathcal{RI_\alpha^Q}(X,Y)$. Recall that $\mathcal{RI_\alpha^Q}(X,Y)$ quantifies the information in comparing the actual distribution $F_1(.)$ with $F_2(.)$. The role is reversed in the case of $\mathcal{RI_\alpha^Q}(Y,X)$.
\end{enumerate}
	\begin{thm}
Let $X$ and $Y$ denote two non-negative rvs. Then,
    \begin{enumerate}
\item $\mathcal{RI_{\alpha}^{Q}}(X,Y) \geq \int_{0}^{1} (1-p)(- Q_3(p))^\alpha q_X(p) dp$ 
\item $\mathcal{RI_{\alpha}^{Q}}(X,Y)\geq A(\alpha)e^{\mathcal{E_Q}(X)}$, where $\mathcal{E_Q}(X)$ is the quantile based differential entropy and $A(\alpha)=e^{\int_{0}^{1}\log[(1-p)(-\log(1-Q_3(p)))^\alpha] dp}$ is a function of $\alpha$.
    \end{enumerate}
	\begin{proof}
  From the well-known inequality $\log x \leq 1-x$, $\forall x \leq 1$, the proof of the theorem's first part readily follows. To obtain the second part, we apply the log-sum inequality,\\
		\begin{eqnarray}\label{4}
			\int_{0}^{1} f(Q(p))\log\frac{f(Q(p))}{(1-p)(-\log(1-Q_3(p)))^\alpha}dQ(p) & \geq & \log \frac{1}{\int_{0}^{1} (1-p) (-\log(1-Q_3(p)))^\alpha q(p)dp} \nonumber\\ 
			& = & -\log (\mathcal{RI}_{\alpha}^{Q}(X,Y)). 
		\end{eqnarray}
		Moreover, LHS of above inequality can be expressed as;
		\begin{equation}\label{5}
			\int_{0}^{1} f(Q(p))\log\frac{f(Q(p))}{(1-p)(-\log(1-Q_3(p)))^\alpha}dQ(p)  = -\mathcal{E_Q}(X) - \int_{0}^{1} \log[(1-p)(-\log(1-Q_3(p)))^\alpha] dp
		\end{equation}
		From \eqref{4} and \eqref{5},
		\begin{equation}\label{6}
			\log (\mathcal{RI_{\alpha}^{Q}}(X)) \geq  \mathcal{E_Q}(X) + \int_{0}^{1} \log[(1-p)(-\log(1-Q_3(p)))^\alpha] dp.   
		\end{equation}
		Exponentiating both sides of \eqref{6} we get,
		\begin{equation*}
			\mathcal{RI_{\alpha}^{Q}}(X) \geq A(\alpha)e^{\mathcal{E}_Q(X)}  
		\end{equation*}
	\end{proof}
    \end{thm}
%\begin{lem}
%\begin{equation}\label{lm2}
%(1-p)(-\log(1-Q_3(p)))^\alpha \leq \int_{0}^{1}[\log  q_1(p)-\log %q_3(p)]^\alpha dp
%\end{equation}
%\begin{proof}
%To prove the result it is enough to establish the following inequality

%\begin{eqnarray}\label{lm2'}
%\bar{F}(x)[-\log \bar{G}(x)]^\alpha & = & \left[1-%\sum_{u=0}^{x}f(u)\right] \left[-\log \left(1-%\sum_{u=0}^{x}g(u)\right)\right] ^\alpha\\
%& \leq & \int_{0}^{\infty}f(x)(-\log g(x))^\alpha dx.
%\end{eqnarray}
%In order to establish the above inequality, we will use a combination of concavity of the logarithmic function, Jensen's inequality and the sum integral inequalities. Since $-\log x$ is a convex function, we can apply Jensen's inequality. The corresponding quantile version can be directly obtained which is given in the lemma. 
%\end{proof}
%\end{lem}
\begin{prop}
   Assume $X$ and $Y$ are rvs having respective QFs, $Q_X$ and $Q_Y$. For $0 \leq \alpha \leq 1$, then the Q-FCRI satisfy the relationship
   \begin{equation*}
      \mathcal{RI_\alpha^Q}(X,Y) \geq [\mathcal{RI^Q}(X,Y)]^\alpha,
   \end{equation*}
   where $\mathcal{RI^Q}(X,Y)=\int_{0}^{1}(1-p)(-\log(1-Q_3(p)))q_X(p)dp$ is the Q-CRI.
   \begin{proof}
  Using $(1-p) \geq [(1-p)]^\alpha$, we have,
  \begin{eqnarray*}
    \mathcal{RI_\alpha^Q}(X,Y)&=&\int_{0}^{1}(1-p)(-\log(1-Q_3(p))^\alpha q_X(p)dp.\\
    {} & \geq & \int_{0}^{1}(1-p)^\alpha[-\log(1-Q_3(p))]^\alpha q_X(p)dp\\
    {}& = & \int_{0}^{1}[-(1-p)\log(1-Q_3(p))q_X(p)]^\alpha dp\\
    {}&= &  \int_{0}^{1} \phi_\alpha(\eta(p))dp
  \end{eqnarray*}
  where, $\eta(p)=-(1-p)\log(1-Q_3(p))q_X(p) \geq 0$ and $\phi_\alpha(z)=z^\alpha$. \\
  
  Here, for $0 \leq \alpha \leq 1$, $\phi_\alpha(z)$ is convex in $z \geq 0$. From the Jensen's inequality, we get
  \begin{equation*}
    \mathcal{RI_\alpha^Q}(X,Y) \geq \bigg[-\int_{0}^{1}(1-p)\log(1-Q_3(p))q_X(p)dp\bigg]^\alpha=[\mathcal{RI^Q}(X,Y)]^\alpha
  \end{equation*}
  Hence, the proof.
   \end{proof}
\end{prop}
\subsection{Q-FCRI of Proportional Hazards Model}
Let $X$ and $Y$ be rvs with hazard functions $h_X(x)$ and $h_Y(x)$, and corresponding SFs $\bar{F}(x)$ and $\bar{G}(x)$. Then $X$ and $Y$ follow the Proportional Hazards Model (PHM) if there exist a positive constant $\beta>0$, such that $h_Y(x)=\beta h_X(x)$, or similarly, $\bar{G}(x)=(\bar{F}(x))^\beta$. The PHM model introduced by \cite{cox1972regression} is widely used in survival analysis, particularly
in medical statistics and reliability engineering, where it captures the effect of covariates (such
as treatment, environment, or subject characteristics) on the lifetime of an individual or system. The FCRI of PHM has the form
\begin{equation}
\mathcal{RI_\alpha} (X,Y) = {\beta}^\alpha \int_{0}^{\infty} \bar{F}(x)(-\log \bar{F}(x))^\alpha dx
\end{equation}
Then, the Q-FCRI of PHM is given by;
\begin{equation}
\mathcal{RI_\alpha^Q} (X,Y) =  \beta^\alpha \int_{0}^{1} (1-p)(-\log(1-p))^\alpha q_X(p)dp=\beta^\alpha \mathcal{E_\alpha^Q}(X).
\end{equation}
\begin{ex}
A series system can be considered as an extended application of PHM. For a  system arranged in series with $n$ independently and identically distributed (iid) components, the Q-FCRI between the iid components and the series system $X_{(1)}$ is
$$\mathcal{RI_\alpha^Q}(X,X_{(1)})=n^\alpha \int_{0}^{1}(1-p)(-\log(1-p))^\alpha q_X(p)dp=n^\alpha \mathcal{E_\alpha^Q}(X),$$
where $\mathcal{E_\alpha^Q}(X)$ is the Q-FCRE given in \eqref{QFCRE}.
\end{ex}

\subsection{Q-FCRI and upper record values}
Assuming a sequence of iid rvs $\{X_n:n \geq 1\}$ with PDF $f(\cdot)$ and CDF $F(\cdot)$. If an observation $X_i$ exceeds all previous values $i.e.$, $X_i > X_j \hspace{0.3cm} \forall \hspace{0.3cm} i>j$, then it is known as an upper record value. Suppose $\{X_m=U_m \}$ denote the $m^{th}$ upper record value, arising from the $\{X_n, n \geq 1\}$, then the survival function $\bar{F}_{U_m}$ and density function $f_{U_m}$ of $U_m$ are given by (see \cite{raju2024results}),
\begin{equation*}
\bar{F}_{U_m}(x)=\sum_{i=0}^{m-1}\frac{[-\log(\bar{F}(x))]^i}{i!}\bar{F}(x)
\end{equation*}
and
\begin{equation*}
f_{U_m}(x)=\frac{[-\log(\bar{F}(x))]^{m-1}}{(m-1)!}f(x), x>0, m\geq1.
\end{equation*}
Then, the FCRI with respect to $U_m$ and the random variable $X$ is 
\begin{eqnarray}
\mathcal{RI_\alpha}(U_m,X)&=&\int_{0}^{\infty} \bar{F}_{U_m}(x)(- \log \bar{F}(x))^\alpha dx\\
{}&=& \int_{0}^{\infty} \bigg(\sum_{i=0}^{m-1}\frac{[-\log(\bar{F}(x))]^i}{i!}\bar{F}(x)\bigg)(-\log \bar{F}(x))^\alpha dx\\
{}&=& \sum_{i=0}^{m-1}\int_{0}^{\infty}\frac{[-\log \bar{F}(x)]^i}{i!}\bar{F}(x)(-\log \bar{F}(x))^\alpha dx\\
{}&=& \sum_{i=0}^{m-1} E_{u_{i+1}}\bigg(\frac{[-\log (\bar{F}(x))]^\alpha}{h_X(x)}\bigg).
\end{eqnarray}
Here, $h_X(X)$ denote the hazard rate function with respect to $F$ and $U_{i+1}$ be the rv with SF $\bar{F}_{U_{i+1}}$. Now, the Q-FCRI corresponding to $U_m$ and the rv $X$ can be expressed as
\begin{equation*}
\mathcal{RI_\alpha^Q}(U_m,X)=\int_{0}^{1} \bigg(\sum_{i=0}^{m-1} \frac{[-\log(1-p)]^i}{i!}(1-p)(-\log(1-p))^\alpha \bigg) q_1(p)dp.
\end{equation*}

The next result provides a tool to calculate Q-FCRI if transformations are applied to $X$ and $Y$.

\begin{thm}\label{thm 2.1}
Let $\tau_1(.)$ and $\tau_2(.)$ be two continuous non-decreasing and invertible transformations, then
\begin{equation}
\mathcal{RI_\alpha^Q}(\tau_1(X), \tau_2(Y))=\int_{0}^{1} (1-p)(- \log(1-Q_2^{-1}(\tau_2^{-1}(\tau_1(Q_1(p))))))^\alpha d \tau_1(Q_1(p)).
\end{equation}
\begin{proof}
Suppose that the survival functions corresponding to $\tau_1(X)$ and $\tau_2(Y)$ be $\bar{F}_{\tau_1(X)}(x)$ and $\bar{F}_{\tau_2(Y)}(x)$ respectively. Then, by using \eqref{fcri} we obtain
\begin{equation}\label{Tqfcri}
\mathcal{RI_\alpha}(\tau_1(X),\tau_2(Y))=\int_{0}^{\infty} \bar{F}_{\tau_1(X)}(x)(-\log \bar{F}_{\tau_2(Y)}(x))^\alpha dx.
\end{equation}
In analogy with \cite{sunoj2018quantile}, the QFs of $F_{\tau_1(X)}(x)$ and $F_{\tau_2(Y)}(x)$ can be shown to be $\tau_1(Q_X(p))$ and $Q_Y^{-1}(\tau_2^{-1}(\tau_1(Q_X(p))))$ respectively. On Substituting these in \eqref{Tqfcri} follows the desired result. 
\end{proof}
\end{thm}

For the illustration of Theorem 2.2, we provide the following examples:

\begin{ex}
Assume $X$ and $Y$ are independent exponentially distributed  rvs with QFs $Q_X(u)=-\frac{1}{\theta_1}\log(1-u)$ and $Q_Y(u)=-\frac{1}{\theta_2}\log(1-u)$, where $\theta_1,\theta_2 > 0$ and $0<u<1$. We know that $\tau_1(X)=X^{\frac{1}{b}}$ and $\tau_2(Y)=Y^{\frac{1}{b}}$, $b>0$, follows Weibull distributions with QFs $\tau_1(Q_X(u))=(-\frac{1}{\theta_1}\log(1-u))^{\frac{1}{b}}$ and $\tau_2(Q_Y(u))=(-\frac{1}{\theta_2}\log(1-u))^\frac{1}{b}$. Then, the QFCRI with respect to $\tau_1(X)$ and $\tau_2(Y)$ is given by
\begin{equation}\label{QFCRI-Webuill}
\mathcal{RI_\alpha^Q}(\tau_{1}(X),\tau_2(Y))=\frac{1}{b}(-1)^{1+\frac{1}{b}+\alpha}\bigg(\frac{\theta_2^\alpha}{\theta_1^{\alpha+\frac{1}{b}}}\bigg) \Gamma\bigg(\frac{1}{b}+\alpha\bigg).
\end{equation}
\end{ex}
The importance of Theorem \ref{thm 2.1} is that the expression for Q-FCRI when both $X$ and $Y$ follow Weibull distribution can be easily determined using \eqref{QFCRI-Webuill}, which on otherwise difficult to compute.
\begin{ex}
It is known if $X$ and $Y$ are two independent rvs following Pareto I distributions with respective QFs $Q_X(u)=(1-u)^{-\frac{1}{a}}$ and $Q_Y(u)=(1-u)^{-\frac{1}{b}}$, then $\tau_1(X)=\log X$ and $\tau_2(Y)=\log Y$ follow exponential densities with corresponding QFs $\tau_1(Q_X(u))=-\frac{1}{a} \log(1-u)$ and $\tau_2(Q_Y(u))=-\frac{1}{b}\log(1-u)$. This implies that, $Q_Y^{-1}(\tau_2^{-1}(\tau_1(Q_X(p))))=1-(1-u)^\frac{b}{a}$. Thus, the QFCRI with respect to $\tau_1(X)$ and $\tau_2(Y)$ is given by
\begin{equation}
\mathcal{RI_\alpha^Q}(\tau_1(X),\tau_2(Y))=\frac{b^\alpha}{a^{\alpha+1}}\Gamma (\alpha+1).
\end{equation}
\end{ex}
\subsection{Stochastic ordering}
Stochastic ordering plays a vital role in probability and statistics as it allows for a comparison of rvs or distributions without the need to fully characterize them. They support analyses with greater robustness and interpretability leading to improved decision making and more effective risk management.\\

We say that $X$ is stochastically lesser than $Y$,  written $X \leq_{st} Y$, whenever $\bar{F}_{X}(x) \leq \bar{F}_{Y}(x) $.
%\begin{prop}
%Assume $Q_{X_1}(.)$, $Q_{X_2}(.)$ and $q_{X_1}$, $q_{X_2}$ represent the QFs and qdfs corresponding to $X_1$ and $X_2$ respectively. Then, if
%\begin{itemize}
%\item [a)] $X_1 \leq_{st} X_2$ and $q_{X_1} < q_{X_2}$, we have $\mathcal{RI_\alpha^Q}(X_1,X_2) \leq min\{\mathcal{RE_\alpha^Q}(X_1), \mathcal{RE_\alpha^Q}(X_2)\}$ 
%\item [b)] $X_1 \geq_{st} X_2$ and $q_{X_1} > q_{X_2}$, we have $\mathcal{RI_\alpha^Q}(X_1,X_2) \geq max\{\mathcal{RE_\alpha^Q}(X_1), \mathcal{RE_\alpha^Q}(X_2)\}$ 
%\end{itemize}
%where $\mathcal{RE_\alpha^Q}(X)=\int_{0}^{1}(1-p)(-\log(1-p))^\alpha q_{X}(p)dp$
%\begin{proof}
%a)Since $X_1 \leq_{st} X_2$, we have $1-p \leq 1-F_{X_2}(Q_1(p)) \implies 1-p \leq 1-Q_3(p) \implies -\log(1-p) \geq -\log(1-Q_3(p))$

%Thus, 
%\begin{eqnarray}\label{st1}
%\int_{0}^{1} (1-p)[-\log(1-p)]^\alpha q_{X_1}(p)dp &\geq& \int_{0}^{1} (1-p)[-\log(1-Q_3(p))]^\alpha q_{X_1}(p)dp\\\nonumber
%\implies \mathcal{RE_\alpha^Q}(X_1) &\geq& \mathcal{RI_\alpha^Q}(X_1,X_2)
%\end{eqnarray}
%Further, under our assumption $q_{X_1} < q_{X_2}$, we get
%\small\begin{eqnarray}\label{st2}
%\int_{0}^{1}(1-p)[-\log(1-Q_3(p))]^\alpha q_{X_1}(p)dp &\leq &\int_{0}^{1} (1-Q_3(p))[-\log(1-Q_3(p))]^\alpha q_{X_2}(p)dp\\\nonumber
%\implies \mathcal{RI_\alpha^Q}(X_1,X_2) &\leq &\mathcal{RE_\alpha^Q}(X_2)\nonumber
%\end{eqnarray}
%On combining the inequalities $\eqref{st1}$ and $\eqref{st2}$, we get $\mathcal{RI_\alpha^Q}(X_1,X_2) \leq min\{\mathcal{RE_\alpha^Q}(X_1), \mathcal{RE_\alpha^Q}(X_2)\}$.
%The proof of part (b) is similar to that of part (a). Hence, it is omitted.
%\end{proof}
%\end{prop}

\begin{prop}
Consider three rvs $X$, $Y$, $Z$ with QFs, $Q_{X}$, $Q_{Y}$, $Q_{Z}$ and QDFs, $q_{X}$, $q_{Y}$, $q_{Z}$ respectively and assume that, $X \leq_{st} Y \leq_{st} \leq Z$. Then,
\begin{itemize}
    \item [a)] $\mathcal{RI_\alpha^Q}(X,Y) \geq \mathcal{RI_\alpha^Q}(X,Z)$
    \item [b)] $\mathcal{RI_\alpha^Q}(Y,X) \geq \mathcal{RI_\alpha^Q}(Y,Z).$
\end{itemize}
\begin{proof}
By the definition of stochastic ordering, $\bar{F}_{X}(x) \leq \bar{F}_{Y}(x) \leq \bar{F}_{Z}(x) \implies  (1-p) \leq 1-Q_{Y}^{-1}(Q_X(p)) \leq 1-Q_Z^{-1}(Q_X(p))$.  Taking, $Q_{Y}^{-1}(Q_X(p))=Q_4(p)$ and $Q_Z^{-1}(Q_X(p))=Q_{5}(p)$, we get the inequality $ (1-p) \leq 1-Q_{4}(p) \leq 1-Q_5(p)$. Thus,
\begin{equation}\label{st3}
(1-p)(-\log(1-Q_4(p)))^\alpha q_X(p) \geq (1-p)(-\log(1-Q_5(p)))^\alpha q_X(p).
\end{equation}
Applying integration on both sides of $\eqref{st3}$ yields part (a) of the proposition. Part (b) of the proposition can be proved similarly.
\end{proof}
\end{prop}

The lemma given below will be useful to show the triangular inequality property of Q-FCRI.
\begin{lem}
    Let the QDFs of $X$, $Y$, $Z$ be denoted by $q_X$, $q_Y$, $q_Z$, respectively. If $X \leq_{st} Y \leq_{st} Z$, then
  \begin{equation*}
    \mathcal{RI_\alpha^Q}(X,Y)+\mathcal{RI_\alpha^Q}(Y,Z) \geq 2 \mathcal{RI_\alpha^Q}(X,Z)
    \end{equation*}
    \begin{proof}
        From the definition of Q-FCRI,
        \begin{eqnarray*}
\mathcal{RI_\alpha^Q}(X,Y)+\mathcal{RI_\alpha^Q}(Y,Z) &=&\int_{0}^{1}(1-p)(-\log(1-Q_4(p)))^\alpha q_X(p)dp \\
{}& + &\int_{0}^{1}(1-Q_4(p))(-\log(1-Q_5(p)))^\alpha q_X(p)dp\\
{} &\geq& \int_{0}^{1} (1-p)(-\log(1-Q_5(p)))^\alpha q_{X}(p)dp\\
{}&+& \int_{0}^{1} (1-p)(-\log(1-Q_5(p)))^\alpha q_{X}(p)dp\\
{}&=& 2 \mathcal{RI_\alpha^Q}(X,Z)
        \end{eqnarray*}
where $Q_4(p)=Q_Y^{-1}(Q_X(p))$, $Q_5(p)=Q_Z^{-1}(Q_X(p))$. Here, the first inequality arises from the assumption $X \leq_{st} Y \leq_{st} Z$. Hence, the proof.
    \end{proof}
\end{lem}
\begin{prop}
Consider rvs $X$, $Y$, $Z$ whose QDFs are $q_X$, $q_Y$, $q_Z$, respectively, satisfying $X \leq_{st} Y \leq_{st} Z$. Then,
\begin{equation*}
 \mathcal{RI_\alpha^Q}(X,Y)+\mathcal{RI_\alpha^Q}(Y,Z) \geq  \mathcal{RI_\alpha^Q}(X,Z).   
\end{equation*}
\begin{proof}
The proof can be straightforwardly obtained from Lemma 2.1. Hence, it can be omitted.
\end{proof}
\end{prop}

\subsection{Q-FCRI based on inverse Mittag-Leffler function}

Recently, researchers (see \cite{di2019past}, \cite{saha2023extended},  \cite{foroghi2023extensions}) explored several properties of fractional versions of uncertainty measures. They found that in complex systems, fractional entropies outperform classical entropies. Fractional uncertainty measures can effectively capture enhanced long-range phenomena, signal evolution sensitivity and non-linear behavior due to well-known features of fractional calculus. \\

Mittag-Leffler \citep{mittag1903nouvelle} showed that the Mittag-Leffler function (MLF) 
 emerges organically during the solution of fractional order integral or
differential equations, and especially in the investigations of the fractional generalization of the kinetic equation, random walks, Lévy flights, superdiffusive transport and in the study of complex systems. The MLF is defined by the following expression
\begin{equation}
E_\alpha(x)=\sum_{r=0}^{\infty}
\frac{x^r}{\Gamma(\alpha r)}, \hspace{0.25cm} 0 < \alpha < 1. 
\end{equation}
%For details of the properties of MLF, one can refer to \cite{haubold2011mittag}. 
We recall that the inverse of the MLF (or the fractional logarithmic function), denoted by $Ln_\alpha(\cdot)$ can be obtained as a solution of (see
 Proposition 7.1 of \cite{jumarie2012derivation})
 \begin{equation}\label{f1}
    f(mn)=f(m)+f(n), \hspace{0.5cm}m,n>0,
 \end{equation}
 where the function $f(\cdot) : \mathbb{R} \rightarrow \mathbb{R}$ has derivative of order $\alpha$, ($0 < \alpha <1$). Moreover, the power of inverse MLF $(Ln_\alpha(\cdot))^\frac{1}{\alpha}$ can be obtained as a solution of
 \begin{equation}\label{f2)}
f^\alpha (mn)=f^\alpha(m)+f^\alpha(n), \hspace{0.2cm} m,n>0,
 \end{equation}
 an extended functional of \eqref{f1}    (see \cite{zhang2020cumulative}).
Various properties of the inverse MLF \citep{jumarie2012derivation} include,
\begin{itemize}
    \item $Ln_\alpha 1=0$, $Ln_\alpha 0=-\infty$, $Ln_\alpha x <0$, when $x<1$;
    \item $1(Ln_\alpha 1)^\frac{1}{\alpha}=0=0(Ln_\alpha 0)^\frac{1}{\alpha}$;
    \item $\frac{d^\alpha}{dx^\alpha}(Ln_\alpha x)^\frac{1}{\alpha}=\frac{\alpha !}{((1-\alpha)!)^2}\frac{1}{x^\alpha}$;
    \item $Ln_\alpha(x^a)=a^\alpha Ln_\alpha x$;
    \item $[Ln_\alpha(mn)]^\frac{1}{\alpha}=[Ln_\alpha(m)]^\frac{1}{\alpha}+[Ln_\alpha(n)]^\frac{1}{\alpha}$.
\end{itemize}
The fractional logarithmic function, denoted by $Ln_\alpha(.)$, has no closed form. Some approximation techniques for computing the inverse MLF are available in the literature (see  \cite{sarumi2020highly}). Motivated by these, we introduce Q-FCRI employing the concept of inverse MLF. \\

 Assume $X$ and $Y$ with SFs $\bar{F}_X$ and $\bar{F}_Y$ respectively
 . Then, for the actual survival function $\bar{F}_X$ and the assigned survival function $\bar{F}_Y$, the FCRI of order $\alpha$ is
 \begin{equation}\label{Mittag-Leffler}
 \mathcal{\widetilde{M}_\alpha}(X,Y)=\int_{0}^{\infty}\bar{F}_X(x)[-Ln_\alpha \bar{F}_Y(x)]^\frac{1}{\alpha} dx = E_{X}\bigg[\frac{[-Ln_\alpha \bar{F}_Y(x)]^\frac{1}{\alpha}}{h_{X}(x)}\bigg], 0 < \alpha<1 
\end{equation}
where and $h_{X}(\cdot)=\frac{f_X(\cdot)}{\bar{F}_X(\cdot)}$ denotes the failure rate of $X$ and $E_{X}(\cdot)$ denotes the expectation with respect to $X$.\\

Since $Ln_\alpha(\cdot)$ has no closed form expression and for computational purposes we use the approximation $Ln_\alpha x \approx \log x^{\alpha !}$ for $0 < \alpha ! <1$. Thus, \eqref{Mittag-Leffler} becomes
\begin{eqnarray*}
\mathcal{\widetilde{M}_\alpha}(X,Y)&=& (\alpha!)^\frac{1}{\alpha} \int_{0}^{\infty}\bar{F}_X(x)[-\log \bar{F}_Y(x)]^\frac{1}{\alpha} dx \\
{}&=& (\alpha !)^\frac{1}{\alpha} E_{X}\bigg[\frac{[-\log \bar{F}_Y(x)]^\frac{1}{\alpha}}{h_{X}(x)}\bigg]
\end{eqnarray*}
provided the integral in RHS is finite.
%Now, the quantile based form of \ref{Mittag-Leffler};
\begin{defn}
Let $Q_X$, $Q_Y$ and $q_X$, $q_Y$ be the QFs and QDFs with respect to $X$ and $Y$ respectively. Then, the quantile-based form of \eqref{Mittag-Leffler} is; 
\begin{eqnarray}\label{Q-ML}
\mathcal{\widetilde{M}_\alpha^Q}(X,Y)&=& \int_{0}^{1} (1-p)(-Ln_\alpha(1-Q_3(p)))^\frac{1}{\alpha} dQ_X(p)\\ \nonumber
{}&=& (\alpha!)^\frac{1}{\alpha} \int_{0}^{1}(1-p)(-\log(1-Q_3(p)))^\frac{1}{\alpha} q_X(p)dp.\nonumber
\end{eqnarray}
\end{defn}
It is crucial to express $\mathcal{\widetilde{M}_\alpha^Q}$ in terms of some well-known measurements. In this context, we introduce the quantile version of the inverse MLF based fractional cumulative residual Kullback-Leibler divergence (KL-divergence) and fractional cumulative residual entropy (FCRE).\\

For $X$ and $Y$ the quantile-based fractional cumulative residual KL-divergence of order $\alpha$  based on inverse MLF is
\begin{equation}\label{MKL}
\mathcal{KL_\alpha^Q}(X,Y)=\int_{0}^{1}(1-p)\bigg[Ln_\alpha \bigg(\frac{1-p}{1-Q_3(p)}\bigg)\bigg]^\frac{1}{\alpha} q_X(p)dp+E(Y)-E(X).
\end{equation}
Similarly, the Q-FCRE of order $\alpha$ based on inverse MLF for $X$ is
\begin{equation}\label{MCE}
\mathcal{CE_\alpha^Q}(X)=\int_{0}^{1}(1-p)(-Ln_\alpha(1-p))^\frac{1}{\alpha} q_X(p)dp.
\end{equation}

The following proposition suggests a relation between the measure we proposed and some well-known measures, such as Q-FCRE and quantile-based fractional cumulative residual KL-divergence.
\begin{prop}
The $\mathcal{\widetilde{M}_\alpha^Q}(X,Y)$ can be expressed as,
\begin{equation*}
\mathcal{\widetilde{M}_\alpha^Q}(X,Y)=\mathcal{CE_\alpha^Q}(X)+(-1)^{\frac{1}{\alpha}+1}\{\mathcal{KL_\alpha^Q}(X,Y)+E(X)-E(Y)\}.
\end{equation*}
\begin{proof}
From \eqref{MCE}, we have
\begin{eqnarray}\label{prop 3.1}
\mathcal{CE_\alpha^Q}(X)&=&\int_{0}^{1}(1-p)(-Ln_\alpha(1-p))^\frac{1}{\alpha} q_X(p)dp\\\nonumber
{}&=& \int_{0}^{1} (1-p)\bigg[-Ln_\alpha \frac{(1-p)}{(1-Q_3(p))} (1-Q_3(p))\bigg]^\frac{1}{\alpha} q_X(p)dp.\nonumber
\end{eqnarray}
By using the result $[Ln_\alpha(mn)]^\frac{1}{\alpha}=[Ln_\alpha(m)]^\frac{1}{\alpha}+[Ln_\alpha(n)]^\frac{1}{\alpha}$, \eqref{prop 3.1} can be rewritten as
\begin{eqnarray*}
 \mathcal{CE_\alpha^Q}(X)&=& \int_{0}^{1} (1-p)[-Ln_\alpha(1-Q_3(p))]^\frac{1}{\alpha} q_X(p)dp + \int_{0}^{1} (1-p)\bigg[-Ln_\alpha\frac{(1-p)}{(1-Q_3(p))}\bigg]^\frac{1}{\alpha} q_X(p)dp \\
 {}&=& \mathcal{\widetilde{M}_\alpha^Q}(X,Y)+(-1)^\frac{1}{\alpha} \{\mathcal{KL_\alpha^Q}(X,Y)-E(Y)+E(X)\}. 
\end{eqnarray*}
Thus, 
\begin{equation*}
\mathcal{\widetilde{M}_\alpha^Q}(X,Y)=\mathcal{CE_\alpha^Q}(X)+(-1)^{\frac{1}{\alpha}+1}\{\mathcal{KL_\alpha^Q}(X,Y)+E(X)-E(Y)\}.
\end{equation*}
Hence, the proof.
\end{proof}
\end{prop}
In probability and statistics, affine transformations are necessary as they condense,  standardize, 
 and preserve the structural and distributional properties of the data. The affine transformation corresponding to rvs $X$ and $Y$ is given as $Y=c X+ d$, $c>0$ and $d \geq 0$. Under an affine transformation, the effect of the measure in \eqref{Q-ML}  is shown in the sequel.

 \begin{prop}
     Let $X$, $Y$, $X'$, $Y'$ denote rvs such that $X'=c X+d$ and $Y'=cY+d$, $c>0$, $d \geq 0$. Then,
     \begin{equation}
        \mathcal{\widetilde{M}_\alpha^Q}(X',Y')=c  \mathcal{\widetilde{M}_\alpha^Q}(X,Y)
     \end{equation}
\begin{proof}
Let $\bar{F}_{X'}(x)=\bar{F}_{X}\bigg(\frac{x-d}{c}\bigg)$ and $\bar{F}_{Y'}(x)=\bar{F}_{Y}\bigg(\frac{x-d}{c}\bigg)$ be the survival functions corresponding to $X'$ and $Y'$ respectively. By the definition of $\mathcal{\widetilde{M}_\alpha}$ we obtain
\begin{eqnarray}\label{affn trans}
\mathcal{\widetilde{M}_\alpha}(X',Y')&=&\int_d^\infty \bar{F}_{X'}(u)[- Ln_\alpha \bar{F}_{Y'}(u)]^\frac{1}{\alpha} du \\ \nonumber
{}&=& \int_d^\infty \bar{F}_{X}\bigg(\frac{u-d}{c}\bigg)\bigg[-Ln_{\alpha} \bar{F}_{Y}\bigg(\frac{u-d}{c}\bigg)\bigg]^\frac{1}{\alpha} du\\ \nonumber
{}&=& c \int_{0}^{\infty} \bar{F}_{X}(u)[-Ln_\alpha \bar{F}_{Y}(u)]^\frac{1}{\alpha} du\\ \nonumber
{}&=& c \mathcal{\widetilde{M}_\alpha}(X,Y).
\end{eqnarray}
Then the corresponding quantile version of \eqref{affn trans} is
\begin{eqnarray}
\mathcal{\widetilde{M}_\alpha^Q}(X',Y')&=&a \int_{0}^{1}(1-p)(-Ln_\alpha(1-Q_3(p)))^\alpha q_X(p)dp\\ \nonumber
{}&=& a \mathcal{\widetilde{M}_\alpha^Q}(X,Y).
\end{eqnarray}
Thus, the proof is completed.
\end{proof}
\end{prop}

\subsection{Q-FCRI for equilibrium distributions}

The equilibrium distributions have a wide range of applications in queuing and reliability theory, since the equilibrium distribution is the limiting distribution of the forward recurrence times in renewal processes. Assuming, $X$ is a non-negative rv with expectation finite, $i.e.,$ $E(X) < \infty$. Then, the PDF corresponding to the equilibrium random variable $X_{e}$ of $X$ is given by (see \cite{gupta2007role}) $$f_{e}(x)=\frac{\bar{F}_X(x)}{E(X)}, \; x>0.$$
We now state the Q-FCRI of a rv and its equilibrium distribution. The quantile-based means of $X$ is denoted as $\mu_{XQ}=\int_{0}^{1}(1-p)q_X(p)dp$ respectively.
\begin{thm}
Suppose $X_e$ be the equilibrium rv corresponding to a rv $X$ with finite expectation. Then, the Q-FCRI between $X$ and $X_e$ is obtained as
\begin{eqnarray*}
\mathcal{RI_\alpha^Q}(X,X_e)&=&\int_{0}^{1}(1-p)\bigg(-\log \frac{1-p}{\mu_{XQ}}\bigg)^\alpha q_X(p)dp\\
{}&=&\int_{0}^{1}[H_{XQ}(p)]^{-1}(\log \mu_{XQ}-\log(1-p))^\alpha dp.
\end{eqnarray*}

\end{thm}
The theorem that follows gives us the Q-FCRI of equilibrium distributions. The quantile-based means corresponding to $X$ and $Y$ is denoted as $\mu_{XQ}=\int_{0}^{1}(1-p)q_X(p)dp$ and $\mu_{YQ}=\int_{0}^{1}Q_X(p)q_3(p)dp$, respectively.
\begin{thm}
Suppose we have equilibrium rvs $X_{e}$ and $Y_{e}$ corresponding to rvs $X$ and $Y$ with finite expectations. Then, the Q-FCRI of $X_{e}$ and $Y_{e}$ is given by
\begin{equation*}
\mathcal{RI_{\alpha}^{Q}}(X_{e},Y_{e})=\int_{0}^{1}\frac{(1-p)}{\mu_{XQ}}\bigg(-\log \frac{1-Q_3(p)}{\mu_{YQ}}\bigg)^\alpha q_X(p)dp.
\end{equation*}
\begin{proof}
The proof can be obtained directly by substitution.
\end{proof}
\end{thm}
When $\alpha=1$, we obtain Theorem 2.2 given in \cite{sankaran2017quantile}. We get $\mathcal{RI_\alpha^Q}(X_{e},Y_{e})=\mathcal{RI_\alpha^Q}(X,Y)$, when $\mu_{XQ}=\mu_{YQ}=1$.

\begin{defn}
Suppose $Q_X(\cdot)$ and $Q_Y(\cdot)$ be the QFs of non-negative absolutely continuous rvs $X$ and $Y$. Then, the quantile-based fractional cumulative residual inaccuracy ratio (Q-FCRIR) is defined by
\begin{equation*}
\mathcal{RIR_\alpha^Q}(X,Y)=\frac{\mathcal{RI_\alpha^Q}(X,Y)
}{\mathcal{E_\alpha^Q}(X)},
\end{equation*}
where $\mathcal{E_\alpha^Q}(X)$ is the quantile based fractional cumulative residual entropy (Q-FCRE).
\end{defn}
$\mathcal{RIR_\alpha^Q}(X,Y)$ is always non-negative. Suppose $X$ and $Y$ are  distributed identically, then, $\mathcal{RIR_\alpha^Q}(X,Y)=1$. Similar to Q-FCRI, Q-FCRIR is also unsymmetric since $\mathcal{RIR}_\alpha^Q(X,Y) \neq \mathcal{RIR}_\alpha^Q(Y,X) $. An explanation on Q-FCRIR can be made like that it quantifies the discrepancy in the information conveyed by the Q-FCRE when the actual SF $\bar{F}_X(\cdot)$ is replaced with assigned SF $\bar{F}_Y(\cdot)$. Moreover, $\mathcal{RIR}_\alpha^Q(X,Y)>1$ implies that the larger information is provided by Q-FCRI rather than Q-FCRE of $F_X(\cdot)$. Some examples of Q-FCRIR are given in the sequel.
\begin{ex}
Let us consider QFs $Q_X(u)=2u-u^2$ and $Q_Y(u)=u$, where $0<u<1$. Then, for $\alpha=0.75$, we have $\mathcal{RI}_\alpha^Q(X,Y)=0.452$ and $\mathcal{E}_\alpha^Q(X)=0.268$. Hence, $\mathcal{RIR}_\alpha^Q=1.682$
\end{ex}
\begin{ex}
Suppose $X$ and $Y$ are distributed exponentially with QFs $Q_X(p)=-\frac{1}{\theta_1}\log(1-p)$ and $Q_Y(p)=-\frac{1}{\theta_2}\log(1-p)$. Then, the Q-FCRIR with respect to $X$ and $Y$ is given by
\begin{equation*}
\mathcal{RIR}_\alpha^Q(X,Y)=\bigg(\frac{\theta_2}{\theta_1}\bigg)^\alpha
\end{equation*}
When $\theta_1=\theta_2$, then Q-FCRIR equals to 1. $\mathcal{RIR}_\alpha^Q(X,Y)>1$ if $\theta_2>\theta_1$ and $\mathcal{RIR}_\alpha^Q(X,Y)<1$ if $\theta_2<\theta_1$.
\end{ex}
\section{Nonparametric estimation of Q-FCRIR}	
Suppose $X_{(1)}, X_{(2)},\ldots, X_{(n)}$ denote the order statistics corresponding to lifetime of $n$ iid components with common CDF $F(x)$ and QF $Q_{X}(u)$. Also, $Y_{(1)}, Y_{(2)},\ldots, Y_{(n)}$ denote the order statistics corresponding to $n$ iid components with common CDF $G(x)$ and QF $Q_{Y}(u)$. Assume the empirical distribution functions of $X$ and $Y$ be $\hat{F}(x_{(i)})$ and $\hat{G}(x_{(i)})$.
\cite{parzen1979nonparametric} introduced an empirical quantile function which is a step function with jump $\frac{1}{n}$ given by
\begin{equation*}
\bar{Q}_{X}(u)=X_{(k)},\hspace{0.2cm}\frac{k-1}{n}<u<\frac{k}{n},\hspace{0.2cm}k=1,2,\ldots,n,
\end{equation*}
 and this estimator's smoothed version is provided by,
\begin{equation*}
\bar{Q}_n(u)=n\bigg(\frac{k}{n}-u\bigg)X_{(k-1)}+n\bigg(u-\frac{k-1}{n}\bigg)X_{(k)}
\end{equation*}
for $\frac{k-1}{n}<u<\frac{k}{n}$, $k=1,2,\ldots,n$. They also defined corresponding estimator of the empirical qdf as
\begin{equation}\label{eqdf}
\bar{q}_n{(u)}=\frac{d}{du}\bar{Q}_n(u)=n(X_{(k)}-X_{(k-1)}), \: \textrm{for} \:\frac{k-1}{n}<u<\frac{k}{n}.
\end{equation}
The empirical plug-in estimator of $Q_3(u)=Q_{Y}^{-1}(Q_{X}(u))$ is
\begin{equation}\label{hat Q_3}
\hat{Q}_3(u)= \hat{G}(\bar{Q}_{X}(u))
\end{equation}
Using \eqref{eqdf} and \eqref{hat Q_3}, the nonparametric estimator of Q-FCRI becomes
\begin{equation}\label{est qfcri}
\hat{\mathcal{RI_\alpha^{Q}}}(X,Y)= \int_{0}^{1}(1-p)(-\log(1-\hat{Q}_3(p)))^\alpha  \bar{q}_n(p)dp  
\end{equation}
Now, by approximating the integral to summation in \eqref{est qfcri}, the plug-in estimator of Q-FCRI is
\begin{equation}\label{est fcrir2}
\hat{\mathcal{RI^Q_\alpha}}(X,Y)=\displaystyle\sum\limits_{i=1}^{n}(1-\hat{F}(X_{(i)}))[-\log(1-\hat{Q}_3(u))]^\alpha n(X_{(i)}-X_{(i-1)})(S_{(i)}-S_{(i-1)})
\end{equation}
where $\hat{F}(X_{(i)})$ is the empirical distribution function of rv $X$ and $S_{(i)}$ is defined by
\[S_{(i)}
=
\begin{cases}
0, i=0  \\
\hat{F}(X_{(i)})=\frac{i}{n}, i=1,2,3,\ldots,n-1 \\
1,i=n
\end{cases}
\]
Hence, the simplified expression of \eqref{est fcrir2} is
\begin{equation}\label{EQFCRI}
\hat{\mathcal{RI_\alpha^{Q}}}(X,Y)
= 
\displaystyle\sum\limits_{i=1}^{n-1}
\left(1-\frac{i}{n}\right)
\left[-\log\left(1-\hat{G}(X_{(i)})\right)\right]^{\alpha}
\, \,(X_{(i)}-X_{(i-1)})\,
\end{equation}
\section{Simulation study}
 This section includes simulations we undertook to assess the performance of the non-parametric estimator $\hat{\mathcal{RI_\alpha^{Q}}}(X,Y)$ in \eqref{EQFCRI}. Initially, we generated pairs of random samples for different sample sizes considering the true model $F$ as power-Pareto distribution having QF $Q_X(u)=Cu^{\lambda_1}(1-u)^{-\lambda_2}, C,\lambda_1,\lambda_2 >0$ which is a quantile model and the assigned distribution $G$ as an exponential distribution with parameter $\theta$. Then we calculate the estimated value of Q-FCRI along with its bias and MSE for $\alpha=0.2$ and $\alpha=0.5$ (see Tables ~\ref{Sim 1} and ~\ref{Sim 2}).
 An increase in the sample size 
$n$ leads to a decrease in both bias and MSE, validating the asymptotic property of the estimator.
\begin{table}[H]
\centering
\caption{Bias and MSE of $\hat{\mathcal{RI_\alpha^{Q}}}(X,Y)$ at $\alpha=0.2$, where $X \sim \text{power-Pareto }   (C=1.5,\lambda_1=0.75,\lambda_2=0.25)$ , $Y \sim \text{Exponential }(\theta=2)$, and the true value is $\mathcal{RI_\alpha^{Q}}(X,Y)=1.3597$.}
\begin{tabular}{@{}p{3.5cm} p{3.5cm} p{3.5cm} p{3cm}@{}} % 4 columns with specified width
	\toprule
	n & $\hat{\mathcal{RI_\alpha^{Q}}}(X,Y)$ & Absolute bias & MSE \\ \midrule
	50 & 1.2305 & 0.1293 & 0.0416 \\
	75 & 1.2610 & 0.0988 & 0.0299\\
	100 &  1.2906 & 0.0691 & 0.0177 \\
	200 &  1.3152 & 0.0446 & 0.0093 \\
	300 & 1.3241 & 0.03561 & 0.0084\\
    500 & 1.3455 & 0.0142 & 0.0032\\
	\bottomrule
\end{tabular}
\label{Sim 1}
\end{table}

\begin{table}[H]
\centering
\caption{Bias and MSE of $\hat{\mathcal{RI_\alpha^{Q}}}(X,Y)$ at $\alpha=0.5$, where $X \sim \text{power-Pareto}(C=1.5,\lambda_1=0.75,\lambda_2=0.25)$ , $Y \sim \text{Exponential }(\theta=2)$, and the true value is $\mathcal{RI_\alpha^{Q}}(X,Y)=1.6367$.}
\begin{tabular}{@{}p{3.5cm} p{3.5cm} p{3.5cm} p{3cm}@{}} % 4 columns with specified width
	\toprule
	n & $\hat{\mathcal{RI_\alpha^{Q}}}(X,Y)$ & Absolute bias & MSE \\ \midrule
	50 & 1.3937 & 0.2429 & 0.1077\\
	75 & 1.4450 & 0.1916 & 0.0789\\
	100 &  1.4929 & 0.1437 & 0.0492 \\
	200 &  1.5338 & 0.1028 & 0.0283 \\
	300 & 1.5536 & 0.0831 & 0.0237\\
    500 & 1.5930 & 0.0436 & 0.0098\\
	\bottomrule
\end{tabular}
\label{Sim 2}
\end{table}

Further performance of the estimator can be validated by generating pairs of random samples for various sample sizes by taking the true model $F$ as Govindarajulu distribution with QF $Q_X(u) = \theta +\sigma((\beta+1)u^{\beta}-\beta u^{\beta+1}),\hspace{0.2cm} \theta \in (-\infty,+\infty), \sigma, \beta > 0, \hspace{0.2cm} 0 \leq u \leq 1$, which is a quantile model and the assigned distribution $G$ as standard uniform distribution. The bias and MSE of $\hat{\mathcal{RI_\alpha^{Q}}}(X,Y)$ relative to the true value are given in Tables ~\ref{Sim 3} and ~\ref{Sim 4}.
\begin{table}[H]
\centering
\caption{Bias and MSE of $\hat{\mathcal{RI_\alpha^{Q}}}(X,Y)$ at $\alpha=0.75$, where $X \sim \text{Govindarajulu}(\theta=0.2,\sigma=0.75,\beta=2)$ , $Y \sim \text{Uniform(0,1) }$, and the true value is $\mathcal{RI_\alpha^{Q}}(X,Y)=0.2915$.}
\begin{tabular}{@{}p{3.5cm} p{3.5cm} p{3.5cm} p{3cm}@{}} % 4 columns with specified width
	\toprule
	n & $\hat{\mathcal{RI_\alpha^{Q}}}(X,Y)$ & Absolute bias & MSE \\ \midrule
	50 & 0.2838 & 0.0077  & 0.0028\\
	75 & 0.2864 & 0.0051 & 0.0016\\
	100 &  0.2877 & 0.0038 & 0.0013 \\
	200 & 0.2890 & 0.0025 & 0.0006 \\
	300 & 0.2905 & 0.0009 & 0.0004\\
    500 & 0.2908 & 0.0006 & 0.0002\\
	\bottomrule
\end{tabular}
\label{Sim 3}
\end{table}
\begin{table}[H]
\centering
\caption{Bias and MSE of $\hat{\mathcal{RI_\alpha^{Q}}}(X,Y)$ at $\alpha=0.85$, where $X \sim \text{Govindarajulu}(\theta=0.2,\sigma=0.75,\beta=2)$ , $Y \sim \text{Uniform(0,1) }$, and the true value is $\mathcal{RI_\alpha^{Q}}(X,Y)=0.2866$.}
\begin{tabular}{@{}p{3.5cm} p{3.5cm} p{3.5cm} p{3cm}@{}} % 4 columns with specified width
	\toprule
	n & $\hat{\mathcal{RI_\alpha^{Q}}}(X,Y)$ & Absolute bias & MSE \\ \midrule
	50 & 0.2797 & 0.0068  & 0.0031\\
	75 & 0.2821 & 0.0045 & 0.0018\\
	100 &  0.2833 & 0.0033 & 0.0014 \\
	200 & 0.2842 & 0.0023 & 0.0007 \\
	300 & 0.2858 & 0.0007 & 0.0004\\
    500 & 0.2860 & 0.0006 & 0.0003\\
	\bottomrule
\end{tabular}
\label{Sim 4}
\end{table}

From Tables ~\ref{Sim 3} and ~\ref{Sim 4}, it can be seen that an increase in sample size results in a reduction of both bias and MSE, which further confirms performance of the estimator $\hat{\mathcal{RI_\alpha^{Q}}}(X,Y)$ based on Govindarajulu (0.2, 0.75, 2) and standard uniform distributions for fractional parameters $\alpha=0.75$ and $\alpha=0.85$. 
\section{Applications}
\subsection{Application to Chaotic Maps}
\subsubsection{Chebyshev map}
"Chebyshev's map" refers to the Chebyshev chaotic map, a simple, efficient, and one-dimensional dynamical system known for its chaotic properties, used in cryptography, image encryption, and biomedical signal analysis. In this subsection, we utilize Chebyshev map discrete-time dynamical system \citep{geisel1984statistical} to generate the artificial time series defined as
\begin{equation*}
x_{i+1}=\cos(a^2 \arccos(x_i)), i=1,\ldots,n-1,
\end{equation*}
where $x_i \in [-1,1]$ and chaotic parameter $a>0$. We initialized the procedure by assigning a starting value $x_1=0.3$ and size $n=1000$.\\

\begin{figure}[H]
\centering
\includegraphics[width=0.75\linewidth]{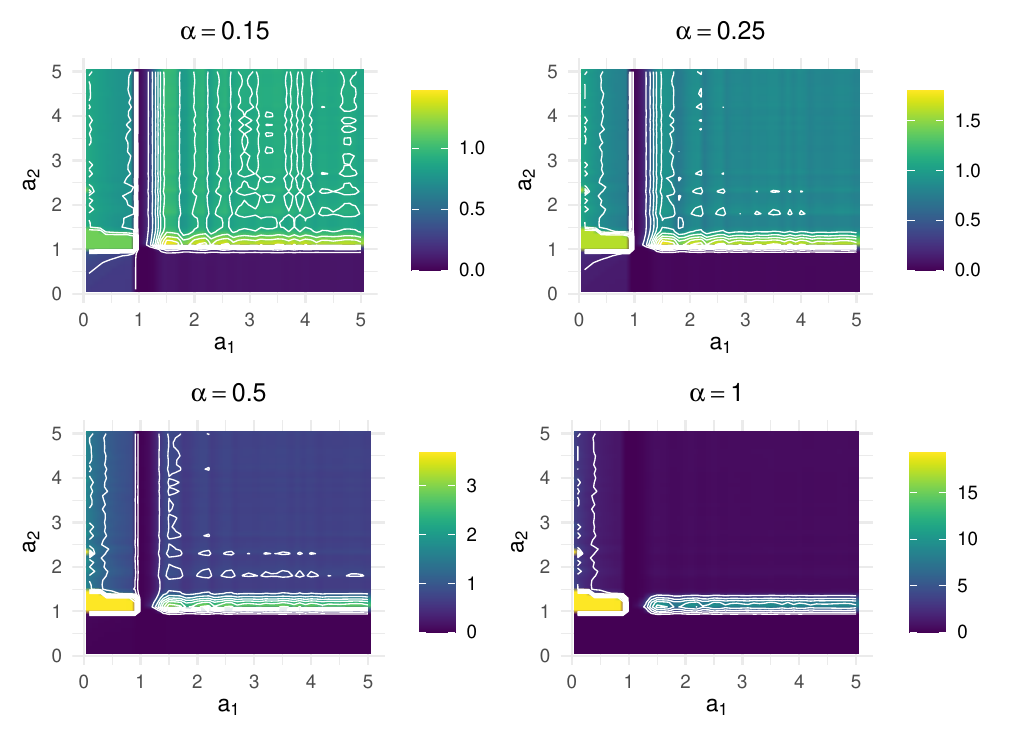}
\caption{Q-FCRI measure between two Chebyshev maps with respective parameters $a_1$ and $a_2$} 
\label{fig1}
\end{figure}

 For $a \geq 1$, we have that $x_i \in [-1,1]$ and, for $0 < a < 1$, $x_i \in [0,1]$ \citep{wang2022pixel}. This range of $a$ is used in estimating the Q-FCRI between two Chebyshev maps with parameters $a_1$ and $a_2$ which has been considered in Figure \ref{fig1}. It is clear that Q-FCRI approximately equals to $0$ for $0< a_2< 1$ and $a_1=1$ where as the Q-FCRI gradually increases for $a_2>1$ and Q-FCRI is highest for $a_2 =1$. It can be visualized that as the fractional parameter $\alpha$ increases, the Q-FCRI value escalates, but its behaviour is similar. However, small values of $\alpha$ enable visualization of dissimilar regions given by $a_1$ and $a_2$.
\subsubsection{Logistic map}
The logistic map discrete-time dynamical system (see \cite{kayal2023weighted}) is defined as
\begin{equation*}
x_{i+1}=cx_i(1-x_i), i=1, \ldots,n-1,
\end{equation*}
where $x_i \in [0,1]$ and $c \in [0,4]$. We initially set value $x_1=0.1$ and size $n=1000$.\\

\begin{figure}[H]
\centering
\includegraphics[width=0.75\linewidth]{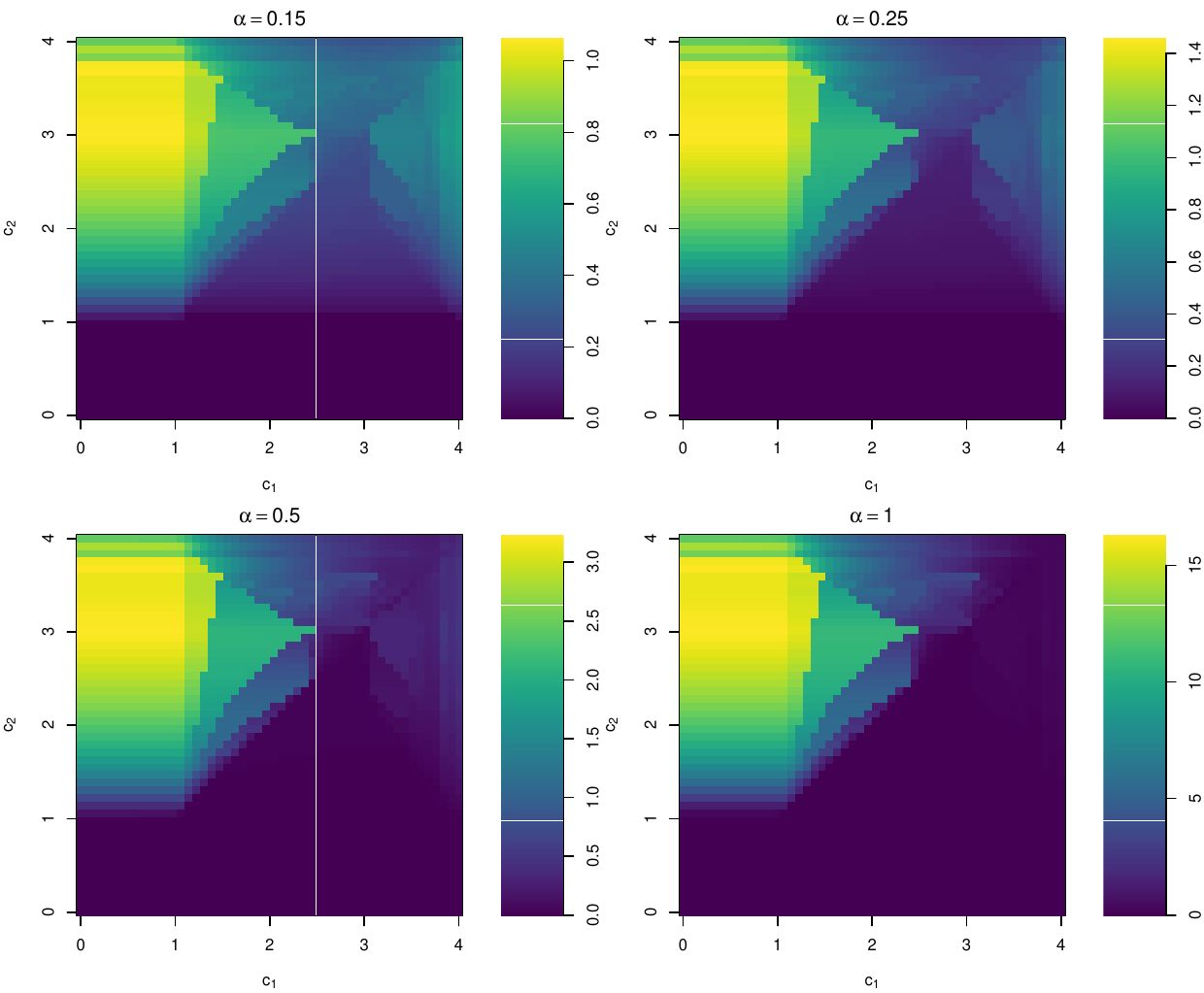}
\caption{Q-FCRI measure for the two Logistic maps characterized by parameters $c_1$ and $c_2$}
\label{fig2}
\end{figure}

\begin{figure}[H]
\centering
\includegraphics[width=0.75\linewidth]{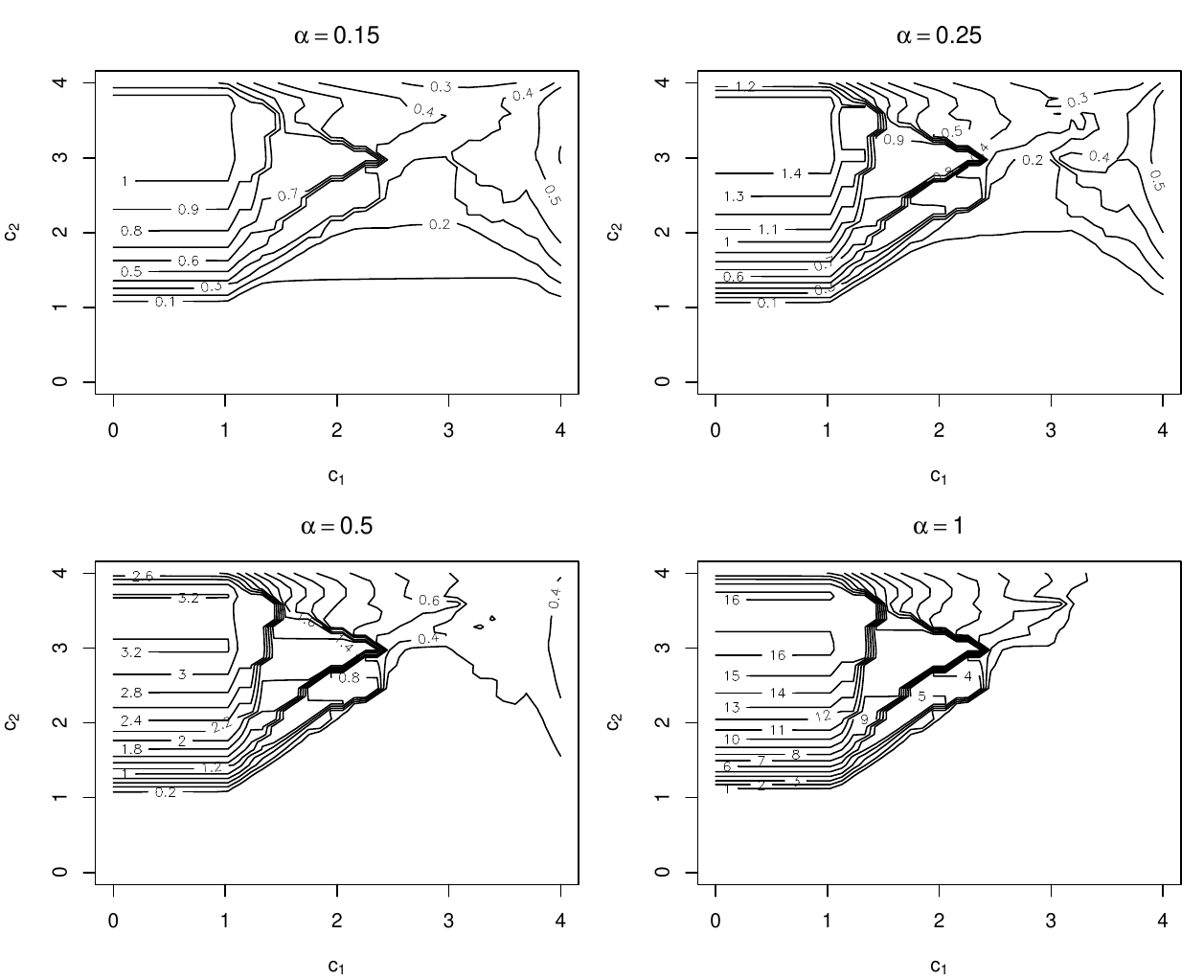}
\caption{Contour line representation of Q-FCRI between two Logistic maps with corresponding parameters $c_1$ and $c_2$}
\label{fig3}
\end{figure}

For estimating the Q-FCRI between the two Logistic map systems, the same interval of c values is employed, which is shown in Figure \ref{fig2}. The Q-FCRI is $0$ in the region $0 < c_2 \leq 1$. However, the discrepancy emerges in $1 < c_2 \leq 4$ and the highest value of Q-FCRI is observed at $2 < c_2 \leq 4$ and $0 \leq c_1 \leq 1$.
The Q-FCRI between the logistic maps depicted by contour lines is shown in Figure \ref{fig3}. Besides, when $\alpha$ increases, the Q-FCRI increases but its behaviour is similar. However, small $\alpha$  values allow in the visualization of discrepancy regions provided by $c_1$ and $c_2$.
\subsection{Detecting Market Uncertainty}
\begin{figure}[H]
\centering
\includegraphics[width=0.55\linewidth]{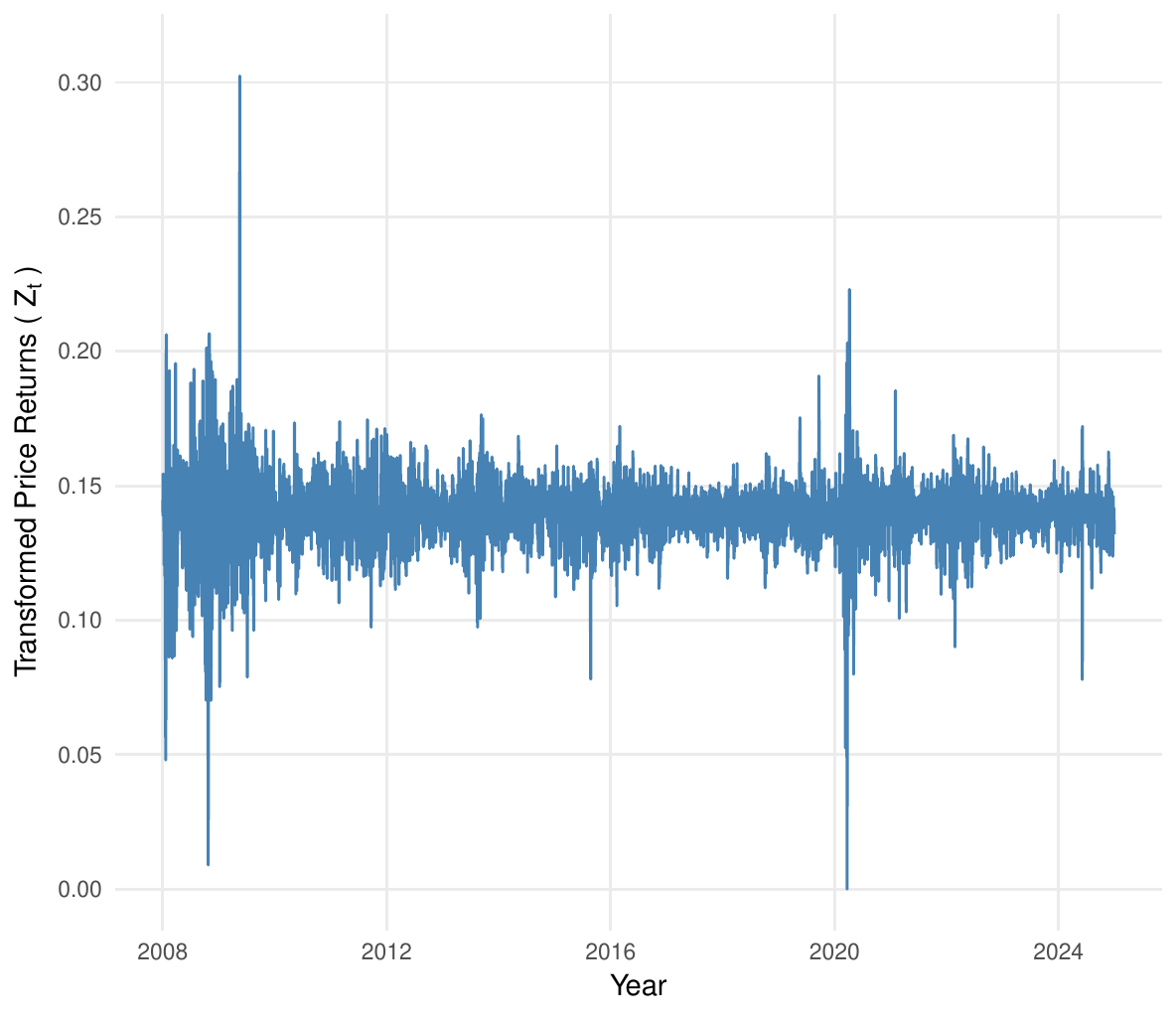}
\caption{Transformed price returns}
\label{fig4}
\end{figure}

\begin{figure}[H]
\centering
\includegraphics[width=0.65\linewidth]{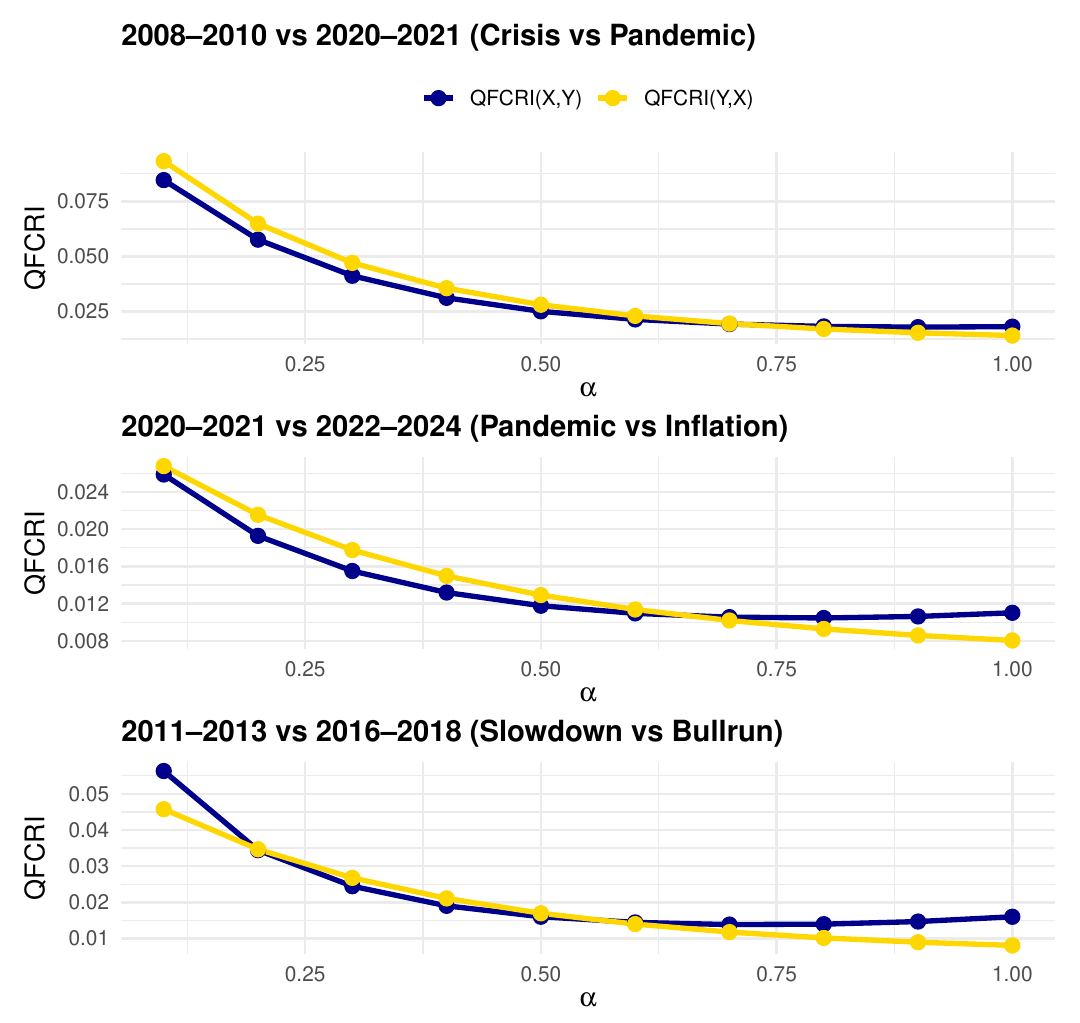}
\caption{Plot of QFCRI at different time regimes of Nifty 50}
\label{fig5}
\end{figure}

We employ the Q-FCRI to measure uncertainty in the Indian stock market by utilizing daily closing prices of the Nifty 50 index from January 1, 2008 to December 31, 2024. Suppose $P_t$ represents the closing price on day $t$ then the daily log returns are calculated as $R_t=\log(P_t)-\log(P_{t-1})$. To make the log returns non-negative, we use the transformation $Z_t=R_t-\min_t R_t$ to get transformed price returns and further calculate the Q-FCRI value (Figure \ref{fig5}). The time series of 
$Z_t$ is depicted in Figure \ref{fig4}, with significant downward spikes indicating episodes of acute market stress.

Figure \ref{fig5} shows three Q-FCRI plots between different time regimes ($i.e.$, crisis v/s pandemic, pandemic v/s inflation, slowdown v/s bullrun). Each plot displays two lines: a blue line for Q-FCRI ($X,Y$) and a yellow line for Q-FCRI ($Y, X$), where $X$ and $Y$ represent the transformed returns of the Nifty 50 index across the respective time periods. The value of Q-FCRI decreases with an increase in $\alpha$ values. Especially when $\alpha=1$; the value of Q-FCRI is very low, suggesting the importance of the fractional parameter $\alpha$ in measuring the discrepancy between financial time regimes. Also, we can see that when $\alpha=1$, the Q-FCRI tends to Q-CRI. Thus, we can easily elucidate that Q-FCRI is more useful than Q-CRI in measuring the discrepancy between two financial time regimes. Also, from Figure \ref{fig5}, the asymmetric property of Q-FCRI, $i.e.$, $\mathcal{RI_\alpha^Q}(X,Y) \neq \mathcal{RI_\alpha^Q}(Y,X)$ can be easily visualised. 
\section{Concluding Remarks}
 Quantile-based Fractional Cumulative Residual Inaccuracy (Q-FCRI) plays a significant role in financial data analysis due to its sensitivity to the fractional parameter $\alpha$
 and its ability to capture the asymmetric discrepancies between financial time regimes.  We have explained an application of Q-FCRI to two chaotic map discrete-time dynamical systems (Chebyshev and Logistic) and demonstrated it to illustrate the behavior of proposed information measures. In a real data application, we employed the Nifty 50 dataset at different time domains, where the value of Q-FCRI decreases as 
$\alpha$ increases, and when $\alpha=1$
, Q-FCRI converges to Q-CRI, demonstrating that the fractional parameter controls the extent and nuance of divergence measured between regimes. Unlike the FCRI in the distributional framework, its quantile framework allows us to use some unique quantile models that do not have any explicit distribution function. The computation of the Q-FCRI for numerical data has been studied by proposing a non-parametric estimator, and the validity of the estimator has also been evaluated using simulation studies. The usefulness of Q-FCRI in analysing the discrepancy/uncertainty between two complex financial time regimes has been examined using Nifty 50 dataset.
\section*{Data Availability Statement}
Data supporting all reported findings in this study are openly accessible at\cite{Yahoo}.

\section*{Research ethics statement}
The data used in this study is obtained from the openly available  \cite{Yahoo}, and ethical approval was not required.

\section*{Conflict of interest statement}
The corresponding author, on behalf of all authors, asserts that there are no conflicts of interest.

\section*{Acknowledgements}
 The first author gratefully acknowledges the financial support provided by the Department of Science and Technology, Government of India, under the INSPIRE Fellowship scheme (Code No: IF220243). 

 \section*{Declaration of generative AI}
 During the preparation of this work, the authors used ChatGPT (OpenAI) to improve the clarity and language of the manuscript. The authors reviewed and edited the content as needed and take full responsibility for the content of the published article.

\bibliographystyle{apalike}
\bibliography{ref} 

\end{document}